\documentclass[conference]{IEEEtran}

\usepackage[T1]{fontenc}
\usepackage[utf8]{inputenc}
\usepackage{amsmath, amssymb, amsfonts}
\usepackage{graphicx}
\usepackage{booktabs}
\usepackage{multirow}
\usepackage{xcolor}
\usepackage{float}
\usepackage{placeins}
\usepackage{textcomp}
\usepackage{array}
\usepackage[hidelinks]{hyperref}
\usepackage{url}
\usepackage{cite}

\begin{document}

\title{A 0.9\,V Bandgap Reference with
7.09\,ppm/\textdegree{}C over a 400\,\textdegree{}C Temperature Range}

\author{
\small
\begin{tabular}{c c c}
\parbox[t]{0.31\textwidth}{\centering
\textbf{Raj Khushal Mali} \\
\textit{Department of Electronics and Computer Science} \\
Vidyalankar Institute of Technology \\
Mumbai, India \\
raj.mali@vit.edu.in \\
ORCID: \href{https://orcid.org/0009-0001-6351-0062}{0009-0001-6351-0062}
}
&
\parbox[t]{0.31\textwidth}{\centering
\textbf{Pooja Pramod Nagvekar} \\
\textit{Department of Electronics and Telecommunication} \\
Vidyalankar Institute of Technology \\
Mumbai, India \\
pooja.nagvekar@vit.edu.in \\
ORCID: \href{https://orcid.org/0009-0002-4816-7419}{0009-0002-4816-7419}
}
&
\parbox[t]{0.31\textwidth}{\centering
\textbf{Dr.\ Akhil Masurkar} \\
\textit{Department of Electronics and Computer Science} \\
Vidyalankar Institute of Technology \\
Mumbai, India \\
akhil.masurkar@vit.edu.in \\
ORCID: \href{https://orcid.org/0000-0001-6797-4086}{0000-0001-6797-4086}
}
\end{tabular}
}

\maketitle

\begin{abstract}
Voltage references used in data converters, power-management ICs, and sensor
interfaces must maintain stable output under wide process, voltage, and
temperature (PVT) variations. While several sub-1\,V CMOS bandgap references
(BGRs) have been reported, their temperature-coefficient (TC) specifications
are usually given over ranges of about 100\,\textdegree{}C to
200\,\textdegree{}C, limiting their applicability in extreme-environment
systems. This paper presents an opamp-based $\beta$-multiplier BGR in GPDK
180\,nm CMOS that achieves, in simulation, \textbf{7.09\,ppm/\textdegree{}C over a
400\,\textdegree{}C range ($-200$\,\textdegree{}C to
$+200$\,\textdegree{}C)}, a span two to four times wider than that of the
prior works compared here.
The proposed circuit combines a PTAT current generator (dual PNP BJTs,
emitter-area ratio $N=8$), a CTAT $V_{BE}$ stage, a two-stage operational
amplifier (65\,dB open-loop gain), and a dedicated start-up circuit, all
implemented and verified in Cadence Virtuoso. Simulation results confirm
$V_{REF} = 0.9$\,V from a 1.8\,V supply, a peak-to-peak variation of only
2.55\,mV, and a temperature derivative of
$-67.16\,\mu$V/\textdegree{}C at 0\,\textdegree{}C, demonstrating the
suitability of the topology for modern low-voltage, wide-temperature analog
and mixed-signal design.
\end{abstract}


\section{Introduction}

Modern electronic systems such as data converters, power management integrated
circuits (PMIC), sensor interfaces, and wireless transceivers rely heavily
on a stable, accurate DC voltage reference as the reference point inside the
circuit against which all other voltages are measured and compared.
The voltage reference block is responsible for generating a DC voltage that
remains invariant under process, supply voltage, and temperature (PVT)
variations~\cite{razavi2001design}. Any drift in this reference directly
affects the performance of dependent blocks, making its design highly critical
in analog IC design.

The silicon bandgap of approximately 1.12\,eV provides a natural voltage scale
that can be exploited to cancel first-order temperature dependence. The
base--emitter voltage $V_{BE}$ of a bipolar junction transistor (BJT) exhibits
a negative temperature dependence (CTAT behaviour), while the thermal voltage
\begin{equation}
V_T = \frac{kT}{q}
\end{equation}
exhibits a positive temperature dependence (PTAT behaviour), where $k$ is
Boltzmann's constant, $T$ is the absolute temperature in Kelvin, and $q$ is
the electron charge. The bandgap reference technique combines these two
complementary dependencies to achieve temperature stability~\cite{widlar1971,brokaw1974,gray2009}.

The output of a first-order compensated BGR is expressed as
\begin{equation}
V_{REF}(T) = V_{BE}(T) + M \cdot V_T(T)
\end{equation}
where the weighting factor $M$ is chosen such that
\begin{equation}
\frac{\partial V_{REF}(T)}{\partial T}
= \frac{\partial V_{BE}(T)}{\partial T}
+ M \cdot \frac{\partial V_T(T)}{\partial T} = 0.
\end{equation}
Since $\partial V_{BE}/\partial T \approx -2\,\text{mV/}^{\circ}\text{C}$ and
$\partial V_T/\partial T = k/q \approx +0.086\,\text{mV/}^{\circ}\text{C}$,
the required $M \approx 23$, yielding $V_{BG} \approx 1.25$\,V, close to
the extrapolated bandgap voltage of silicon at 0\,K.

Progressive CMOS scaling imposes stringent constraints on supply voltages.
At the 180\,nm node, $V_{DD}$ commonly falls to 1.8\,V or lower, making a
1.25\,V reference unsuitable. Generating a sub-1\,V reference while preserving
low TC requires current-mode resistive division~\cite{banba1999,leung2002},
threshold-voltage based references~\cite{ueno2009}, or sub-threshold MOSFET
operation~\cite{devita2007,magnelli2011}, each with trade-offs in stability,
power, and area.

\subsection*{Research Gap}

Despite the progress in sub-1\,V BGR design, a limitation persists
in the published literature: low TC is usually reported over temperature
ranges of about 100\,\textdegree{}C to 200\,\textdegree{}C.
For instance, Banba \textit{et al.}~\cite{banba1999} characterise their
circuit from $27$ to $+125$\,\textdegree{}C (98\,\textdegree{}C span);
Ueno \textit{et al.}~\cite{ueno2009} evaluate their design from $-20$ to
$+80$\,\textdegree{}C (100\,\textdegree{}C span); and even a reference designed
specifically for a wide temperature range~\cite{andreou2012} is characterised
over at most 200\,\textdegree{}C. Applications in automotive,
avionics, deep-space instrumentation, and industrial control require stable
references across ranges of 300\,\textdegree{}C or more, yet none of the
sub-1\,V references surveyed here has been characterised over such a wide
window. This gap motivates the proposed design.

\subsection*{Contributions}

The main contributions of this work are:
\begin{enumerate}
  \setlength{\itemsep}{1pt}
  \setlength{\parskip}{0pt}
  \setlength{\topsep}{2pt}
  \item A 0.9\,V BGR in GPDK 180\,nm CMOS achieving
        \textbf{7.09\,ppm/\textdegree{}C} in simulation over
        $-200$\,\textdegree{}C to $+200$\,\textdegree{}C, a temperature span
        two to four times wider than that of the prior works compared in
        Table~\ref{tab:perf}.
  \item A custom two-stage operational amplifier with 65\,dB open-loop gain
        designed and characterised for precise $V_A = V_B$ enforcement across
        the full temperature range.
  \item Comprehensive Cadence Virtuoso verification including DC, temperature
        sweep, AC and transient analyses, with benchmarking against
        representative prior works.
\end{enumerate}

Unlike curvature-correction techniques~\cite{malcovati2001,leung2003,andreou2012}
that require additional active circuitry, the proposed topology
achieves sub-10\,ppm/\textdegree{}C performance through
precise resistor ratioing alone, keeping the design
compact and compatible with standard GPDK 180\,nm
process design rules without any process modifications.

\vspace{-4pt}
\section{Literature Survey}

\vspace{-2pt}
\subsection{PTAT and CTAT Voltage Generation}

The temperature behaviour of the base--emitter voltage of a BJT biased at
collector current $I_C$ is governed by
\begin{equation}
  V_{BE}(T) = V_{G0}\!\left(1 - \frac{T}{T_0}\right)
              + V_{BE0}\frac{T}{T_0}
              - \frac{\eta k T}{q}\ln\!\frac{T}{T_0},
  \label{eq:VBE_T}
\end{equation}
where $V_{G0} \approx 1.205$\,V, $T_0$ is a reference temperature,
$V_{BE0}$ is $V_{BE}$ at $T_0$, and $\eta \approx 3.5$--$4$ is a
process-dependent constant~\cite{gray2009}. The dominant linear term yields
$\partial V_{BE}/\partial T \approx -2$\,mV/\textdegree{}C, confirming the
CTAT nature of $V_{BE}$.

A PTAT voltage is derived from the difference in base--emitter voltages of two
BJTs ($Q_1$, $Q_2$) biased at different current densities with emitter-area
ratio $N$:
\begin{equation}
  \Delta V_{BE} = V_{T}\ln N.
  \label{eq:DVBE}
\end{equation}
Since $V_T = kT/q$, $\Delta V_{BE}$ rises linearly with absolute temperature,
providing the PTAT component required for TC compensation.

\subsection{Widlar Bandgap Voltage Reference}

The Widlar BGR~\cite{widlar1971} achieves first-order TC cancellation via the
weighted sum
\begin{equation}
  V_{REF}(T) = m_1 V_{PTAT}(T) + m_2 V_{CTAT}(T),
  \label{eq:Widlar_sum}
\end{equation}
with $m_1$, $m_2$ satisfying
\begin{equation}
  m_1 \frac{\partial V_{PTAT}}{\partial T}
  + m_2 \frac{\partial V_{CTAT}}{\partial T} = 0.
  \label{eq:Widlar_TC}
\end{equation}
Because cancellation is exact only at $T_{nom}$, higher-order curvature terms
in $V_{BE}(T)$ introduce residual TC at extreme temperatures, motivating
curvature-correction techniques for wide-range
designs~\cite{malcovati2001,leung2003,andreou2012}.

\subsection{Sub-1\,V Bandgap Voltage Reference}

Three principal strategies lower the reference voltage below 1\,V.

\subsubsection{Resistive Division}
\begin{equation}
  V_{REF,low} = \frac{R_2}{R_1 + R_2} \cdot V_{BG}.
  \label{eq:res_div}
\end{equation}
In practice the division is done in the current domain: the PTAT and CTAT
currents are summed and converted to a voltage by a single
resistor~\cite{banba1999,leung2002}. This is straightforward to implement,
but increases chip area and can degrade noise performance. Leung and
Mok~\cite{leung2002} report 15\,ppm/\textdegree{}C over 0 to
100\,\textdegree{}C with a minimum supply of 0.98\,V. The TC performance
degrades at extreme temperatures where resistor mismatch becomes significant.

\subsubsection{Threshold-Voltage Based References}
MOSFET threshold voltages decrease with temperature and can be combined with a
PTAT voltage to give a compact sub-1\,V reference. Ueno
\textit{et al.}~\cite{ueno2009} generate the extrapolated threshold voltage of
a MOSFET at 0\,K, about 745\,mV, using no resistors, and report
15\,ppm/\textdegree{}C on average (7\,ppm/\textdegree{}C at best) over $-20$
to $+80$\,\textdegree{}C at 0.3\,$\mu$W in 0.35\,$\mu$m CMOS. The output
depends directly on the process threshold voltage, and the specified range is
limited to 100\,\textdegree{}C.

\subsubsection{Sub-threshold MOSFET Operation}
In weak inversion the drain current follows
\begin{equation}
  I_D = I_0 \exp\!\left(\frac{V_{GS} - V_{th}}{nV_T}\right),
  \label{eq:subthreshold}
\end{equation}
enabling purely CMOS PTAT generators without explicit
BJTs~\cite{devita2007,magnelli2011,zhuang2014,wang2015}. Such references reach
nanowatt power, for example 2.6\,nW at 0.45\,V~\cite{magnelli2011} and
14.6\,nW at 0.45\,V~\cite{wang2015}, but they are usually characterised over
narrow temperature ranges, making them less suited to wide-temperature
applications.

\subsection{Gap Analysis and Motivation for This Work}

A critical examination of the prior art reveals a consistent shortcoming:
TC values below 20\,ppm/\textdegree{}C are reported over temperature ranges
of 100\,\textdegree{}C to 200\,\textdegree{}C. Current-mode resistive-division
references~\cite{banba1999,leung2002} are characterised over about
100\,\textdegree{}C; the threshold-voltage reference of~\cite{ueno2009} over
100\,\textdegree{}C; and the wide-range curvature-compensated reference
of~\cite{andreou2012} reports 3.9\,ppm/\textdegree{}C over 165\,\textdegree{}C
and 13.7\,ppm/\textdegree{}C over 200\,\textdegree{}C. The trade-off between
TC and temperature range is the primary unaddressed challenge. The proposed
opamp-based $\beta$-multiplier addresses this gap by combining precision
feedback-enforced current balancing with a carefully ratioed resistor network,
achieving 7.09\,ppm/\textdegree{}C over a 400\,\textdegree{}C span in
simulation. Furthermore, the proposed design operates from a standard
1.8\,V supply without requiring special low-threshold devices
or process modifications, making it directly suitable for
system-on-chip integration in mainstream 180\,nm CMOS processes.

\section{Proposed Circuit Design}

The proposed BGR operates from a 1.8\,V supply and produces
$V_{REF} \approx 0.9$\,V. It comprises four primary sub-blocks: a start-up
circuit, a PTAT current generator, a CTAT voltage stage, and an opamp-based
summation network, described in detail below.

\subsection{Overall Architecture}

\begin{figure}[H]
  \centering
    \includegraphics[width=\columnwidth]{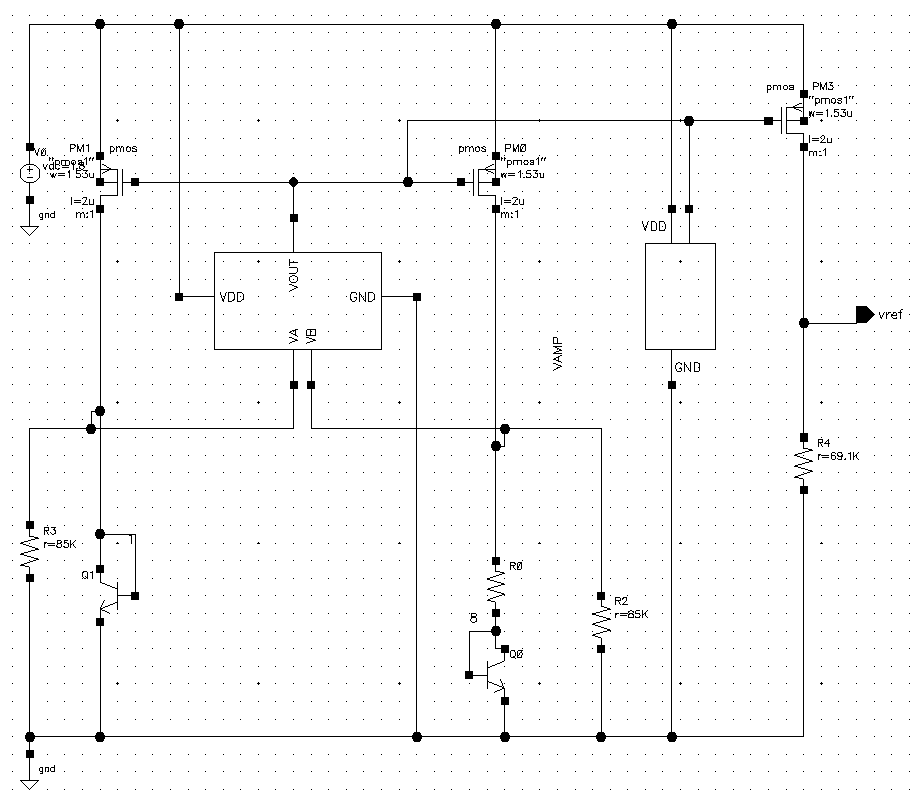}
  \caption{Complete proposed BGR circuit schematic (Cadence instance names:
           $PM_1$, $PM_0$, $PM_3$ correspond to $M_1$, $M_2$ and the output
           mirror; $Q_0$ ($\times8$) and $R_0$ correspond to $Q_2$ and $R_1$
           of Fig.~\ref{fig:bgr_core}).}
  \label{fig:main_sche}
\end{figure}

Figure~\ref{fig:main_sche} shows the complete circuit topology of the proposed
BGR. The start-up circuit initialises the bias network at power-on by injecting
a transient current; once the core reaches its correct operating point, the
start-up path becomes inactive. The PTAT current generator exploits
$\Delta V_{BE}$ of two substrate PNP BJTs ($Q_1$, $Q_2$) with emitter-area
ratio $N$ to produce a current that rises linearly with absolute temperature.
A PMOS current mirror ($M_1$, $M_2$) replicates this current into the output
branch through a third, identical PMOS device. The CTAT voltage stage is formed
by the $V_{BE}$ of diode-connected $Q_1$, whose negative temperature slope
counteracts the PTAT component. The
operational amplifier closes the feedback loop by forcing $V_A = V_B$,
ensuring accurate current splitting independent of temperature and supply.
The summation and output stage combines the PTAT current and CTAT voltage
through the resistor network ($R_1$--$R_4$); the ratio $R_4/R_3$ provides
resistive division that scales $V_{REF}$ to the target 0.9\,V.

\subsection{Sub-1\,V BGR Topology: Opamp-Based $\beta$-Multiplier}

\subsubsection{Topology Selection Rationale}

Among the three sub-1\,V strategies reviewed in Section~II, the
opamp-based $\beta$-multiplier (current-sum) topology was selected for
the following reasons. First, current-mode resistive
division~\cite{banba1999,leung2002} relies on matched resistor ratios; the
opamp feedback loop in the $\beta$-multiplier enforces $V_A = V_B$
independent of absolute resistor values, making it more robust to
process variation at extreme temperatures. Second, threshold-voltage
based references~\cite{ueno2009} set the output by the process threshold
voltage and are specified over only 100\,\textdegree{}C.
Third, sub-threshold MOSFET references~\cite{devita2007,magnelli2011} achieve
ultra-low power but are characterised over narrow temperature ranges. The
$\beta$-multiplier combines the precision of BJT-based $V_{BE}$ references
with opamp-enforced symmetry, enabling sub-10\,ppm/\textdegree{}C
performance over a much wider temperature range than the sub-1\,V works
compared in this paper.

\subsubsection{Operating Principle}

The BGR topology is a modified opamp-based $\beta$-multiplier circuit adapted
to achieve 900\,mV output, drawing upon the principle of~\cite{banba1999},
itself an adaptation of the classical Brokaw cell~\cite{brokaw1974}. Rather
than stacking two thermally complementary voltages in series, the circuit
operates as a \textit{current-sum} topology: a temperature-independent current
is converted to a reference voltage through a precision output resistor.

With $R_2 = R_3$, so that $I_1 = I_2$, $I_{1a} = I_{2a}$, and
$I_{1b} = I_{2b}$, the individual branch currents are:
\begin{equation}
  I_{2a} = \frac{V_B}{R_3} = \frac{V_A}{R_3} = \frac{V_{BE1}}{R_3},
  \label{eq:I2a}
\end{equation}
\begin{equation}
  I_{2b} = \frac{V_B - V_{BE2}}{R_1}
         = \frac{V_{BE1} - V_{BE2}}{R_1}
         = \frac{\Delta V_{BE1,2}}{R_1},
  \label{eq:I2b}
\end{equation}
where $\Delta V_{BE1,2} = V_T \ln(N)$ from~\eqref{eq:DVBE}. The mirrored
current $I_{2a} + I_{2b}$ flows through $R_4$, and the output voltage
simplifies to:
\begin{equation}
  \boxed{V_{REF} = \frac{R_4}{R_3}
            \left(V_{BE1} + \frac{R_3}{R_1}\,V_T \ln N\right),}
  \label{eq:VREF_opamp}
\end{equation}
confirming first-order TC compensation with weighting factor
$M = (R_3/R_1)\ln(N)$, and resistive scaling $R_4/R_3$ yielding
$V_{REF} \approx 0.9$\,V.

\subsection{BGR Core Design}

\begin{figure}[H]
  \centering
    \includegraphics[width=\columnwidth]{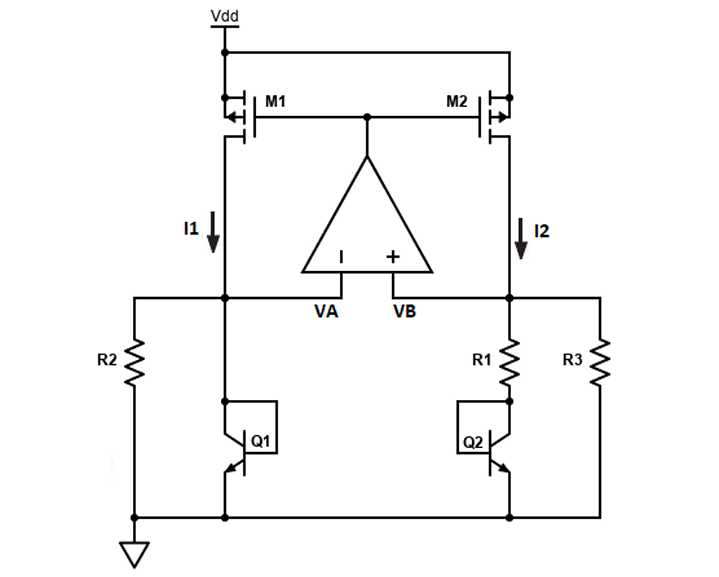}
  \caption{BGR core topology.}
  \label{fig:bgr_core}
\end{figure}

The BGR core consists of two substrate PNP BJTs ($Q_1$, $Q_2$), a PMOS
current mirror ($M_1$, $M_2$), an operational amplifier, and the resistor
network ($R_1$, $R_2$, $R_3$). Figure~\ref{fig:bgr_core} shows the BGR core
topology.

\subsubsection{Initial Bias Condition and Current Setting}

\begin{figure}[H]
  \centering
    \includegraphics[width=\columnwidth]{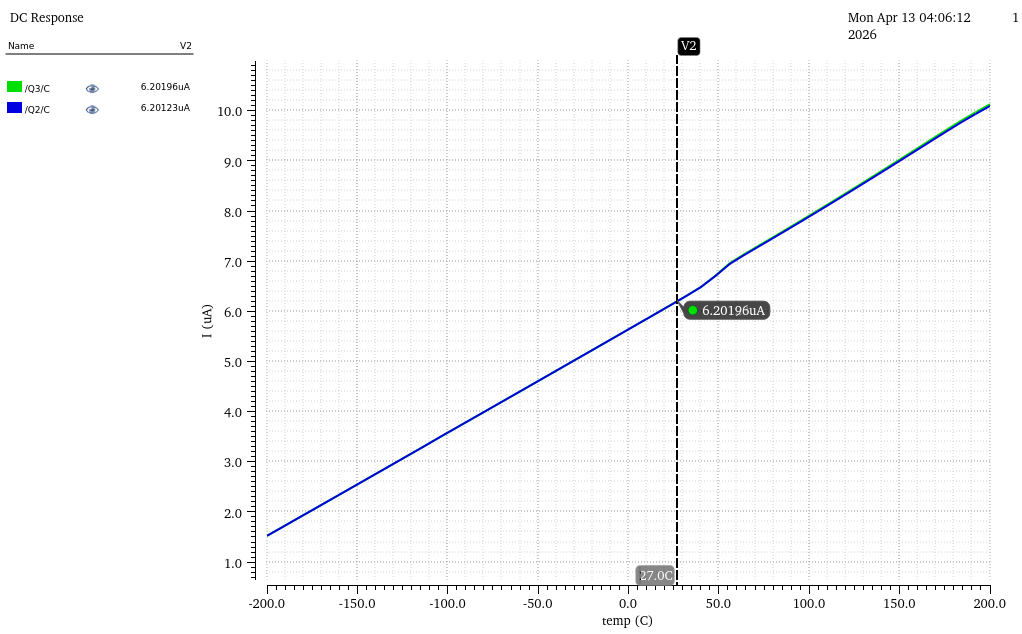}
  \caption{Simulated collector currents of the two BJTs vs.\ temperature
           (Cadence instances $Q_3$ and $Q_2$), confirming equal bias
           currents at 27\,\textdegree{}C.}
  \label{fig:IC_equal}
\end{figure}

The bipolar transistors $Q_1$ and $Q_2$ must be biased in the forward-active
region to establish well-defined $V_{BE}$ values. The quiescent collector
current at $T_{nom} = 27$\,\textdegree{}C is set as:
\begin{equation}
  I_{C1} = I_{C2} = 6.20196\;\mu\text{A}.
  \label{eq:Ibias}
\end{equation}
Figure~\ref{fig:IC_equal} confirms equal collector currents at $T_{nom}$
(6.20196\,$\mu$A and 6.20123\,$\mu$A) and shows that both currents rise
linearly with temperature, as expected of a PTAT current.

\subsubsection{Virtual Short-Circuit Condition}

\begin{figure}[H]
  \centering
    \includegraphics[width=\columnwidth]{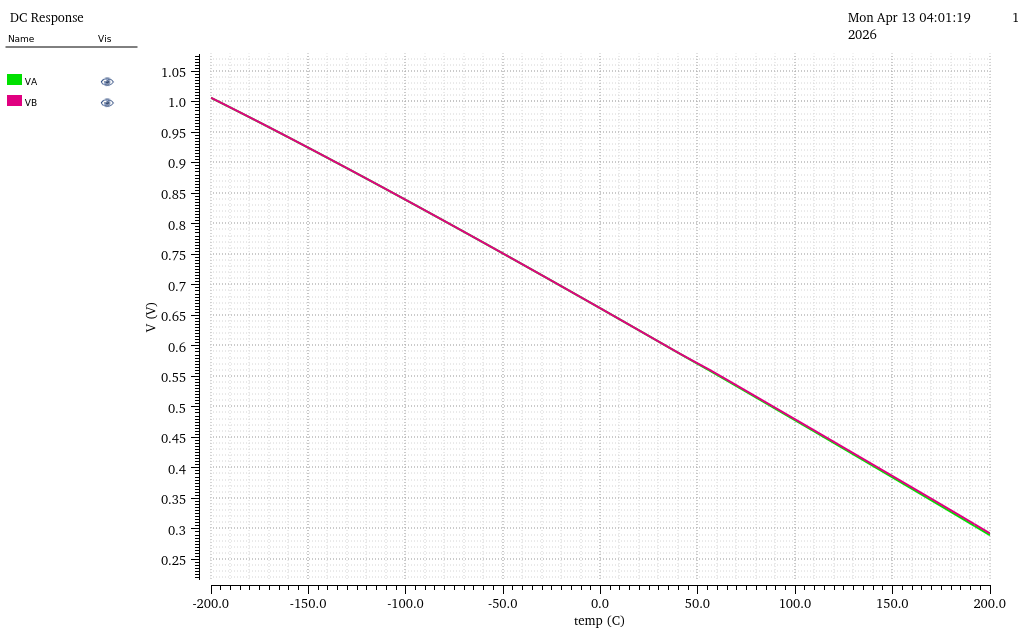}
  \caption{Simulated $V_A$ and $V_B$ vs.\ temperature, confirming the
           virtual short-circuit condition.}
  \label{fig:VA_VB}
\end{figure}

The opamp enforces $V_A = V_B$ through negative feedback. With $R_2 = R_3$,
this symmetry forces $I_1 = I_2$ in both branches. The node $V_B$ is clamped
to $V_{BE1}$, so the voltage across $R_1$ equals:
\begin{equation}
  V_{R_1} = V_{BE1} - V_{BE2} = \Delta V_{BE1,2},
  \label{eq:VR1}
\end{equation}
yielding the pure PTAT current $I_{PTAT} = \Delta V_{BE1,2}/R_1$.
Figure~\ref{fig:VA_VB} confirms $V_A = V_B$ across the temperature sweep;
both nodes follow the CTAT slope of $V_{BE1}$, from about 1.0\,V at
$-200$\,\textdegree{}C to about 0.29\,V at $+200$\,\textdegree{}C.

\subsubsection{\texorpdfstring{$\Delta V_{BE1,2}$}{Delta VBE} Extraction}

\begin{figure}[H]
  \centering
    \includegraphics[width=\columnwidth]{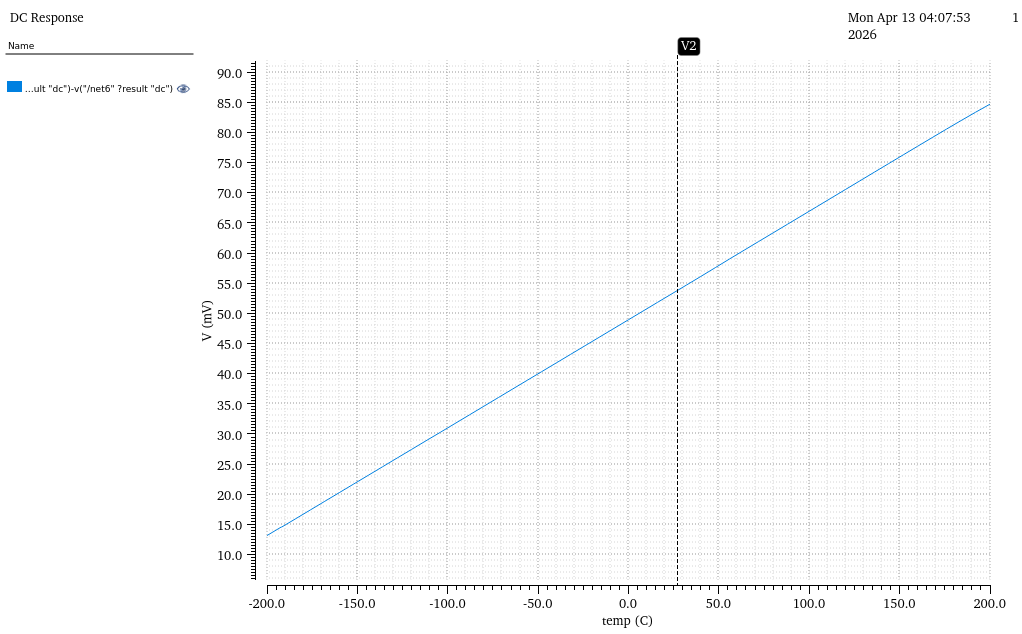}
  \caption{Simulated $\Delta V_{BE1,2}$ vs.\ temperature.
           At $T_{nom} = 27$\,\textdegree{}C: $\Delta V_{BE1,2} \approx 53.78$\,mV.}
  \label{fig:dVBE12_plot}
\end{figure}

$Q_1$ (unit area, $1\times$) and $Q_2$ ($N = 8\times$) are biased at equal
currents $I_{C1} = I_{C2}$. The larger area of $Q_2$ reduces its current
density, giving a smaller $V_{BE2}$:
\begin{equation}
  \Delta V_{BE1,2} = V_{BE1} - V_{BE2} = V_T \ln N.
  \label{eq:DVBE12_detail}
\end{equation}
At 27\,\textdegree{}C, $V_T \ln 8 = 25.86\,\text{mV} \times 2.0794
= 53.77$\,mV, in agreement with the simulated value of
Fig.~\ref{fig:dVBE12_plot}.

\subsubsection{Resistor Sizing}

From simulation at $T_{nom}$:
$\partial V_{BE1}/\partial T = -1.837$\,mV/\textdegree{}C and
$\partial V_T/\partial T \approx +0.09$\,mV/\textdegree{}C,
giving $M \approx 1.837/0.09 \approx 20.4$. With $\Delta V_{BE1,2} = 53.78$\,mV:
\begin{equation}
  R_1 = \frac{53.78\;\text{mV}}{6.20196\;\mu\text{A}} = 8.671\;\text{k}\Omega,
  \label{eq:resistors}
\end{equation}
\begin{equation}
  R_2 = R_3 = \frac{M R_1}{\ln N} = \frac{20.4 \times 8.671\;\text{k}\Omega}{2.079}
  \approx 85\;\text{k}\Omega .
\end{equation}
The output resistor is $R_4 = 69.1$\,k$\Omega$, so that $R_4/R_3 = 0.813$
scales the output to 0.9\,V.

\subsubsection{PMOS Current Mirror Sizing}
$M_1$, $M_2$ and the output mirror device are sized for $V_{GS} = 1.16$\,V:
$W = 1.53\,\mu$m, $L = 2\,\mu$m (Table~\ref{tab:mosfet}).

\begin{table}[H]
\centering
\caption{PMOS Current Mirror Transistor Dimensions}
\label{tab:mosfet}
\renewcommand{\arraystretch}{1.3}
\begin{tabular}{lcc}
\toprule
\textbf{Component} & \textbf{W\,(\textmu m)} & \textbf{L\,(\textmu m)} \\
\midrule
$M_1$ ($PM_1$) & 1.53 & 2.0 \\
$M_2$ ($PM_0$) & 1.53 & 2.0 \\
Output mirror ($PM_3$) & 1.53 & 2.0 \\
\bottomrule
\end{tabular}
\end{table}

\subsubsection{Transistor Sizing Procedure}

\begin{figure}[H]
  \centering
    \includegraphics[width=\columnwidth]{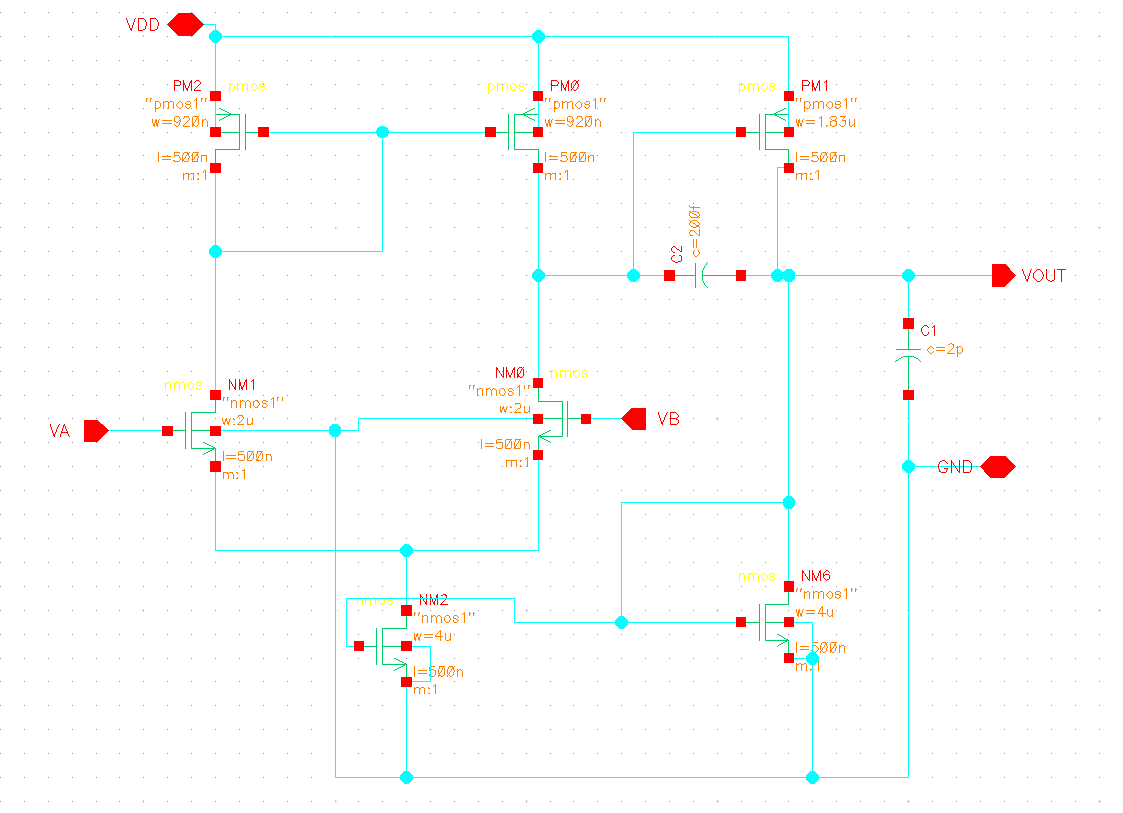}
  \caption{Two-stage opamp schematic.}
  \label{fig:opamp_schem}
\end{figure}

Each opamp transistor width is determined using Cadence DC $I_D$--$W$ sweeps
at the target bias point, with $L = 500\,\text{nm}$ for all devices.
The overdrive voltage for the NMOS current-source devices is set to
$5\%$ of $V_{DD}$, i.e.\
$V_{GS}-V_{TH} = 0.05\times1.8\,\text{V} = 0.09\,\text{V}$,
giving a gate bias of $V_G = V_S + V_{TH} + 0.09\,\text{V}$.
Table~\ref{tab:opamp_sizing} lists the devices with their Cadence instance
names (Fig.~\ref{fig:opamp_schem}).

\begin{table}[H]
\centering
\caption{Two-stage op-amp transistor sizing summary
         (GPDK\,180; $L = 500\,\text{nm}$ for all devices).}
\label{tab:opamp_sizing}
\renewcommand{\arraystretch}{1.25}
\begin{tabular}{@{}lllcc@{}}
\toprule
\textbf{Device} & \textbf{Type} & \textbf{Role}
  & $W\ (\mu\text{m})$ & $I_D\ (\mu\text{A})$ \\
\midrule
$NM_1$, $NM_0$ & NMOS & Diff.\ input pair        & 2.00 & 10 \\
$PM_2$, $PM_0$ & PMOS & Current-mirror load       & 0.92 & 10 \\
$NM_2$         & NMOS & Stage-1 tail source       & 4.00 & 20 \\
$PM_1$         & PMOS & CS output stage           & 1.83 & 20 \\
$NM_6$         & NMOS & Stage-2 current source    & 4.00 & 20 \\
\bottomrule
\end{tabular}
\end{table}

The sizing rationale for each device is given below.

\medskip
\noindent\textbf{$NM_2$ --- Stage-1 tail current source
  (NMOS, $I_D = 20\,\mu$A).}\;
Setting $V_{ov} = 5\%\,V_{DD}$:
\begin{equation}
  V_{GS}-V_{TH} \;=\; 0.05\times1.8\,\text{V} \;=\; 0.09\,\text{V},
\end{equation}
\begin{equation}
  V_{G,NM_2} \;=\; V_S + V_{TH} + 0.09 \;=\; 0.590\,\text{V}.
\end{equation}
The Cadence sweep gives $W = 3.99\,\mu\text{m}$, rounded to $4\,\mu$m in the
schematic.

\medskip
\noindent\textbf{$NM_1$, $NM_0$ --- Differential input pair
  (NMOS, $I_D = 10\,\mu$A each).}\;
Sized at the edge of saturation ($V_{GS}-V_{TH}=V_{DS}$):
\begin{equation}
  V_G \;=\; V_D + V_{TH} \;=\; 0.9 + 0.5 \;=\; 1.4\,\text{V}.
\end{equation}
The Cadence sweep gives $W = 2.00\,\mu\text{m}$.

\medskip
\noindent\textbf{$NM_6$ --- Stage-2 current source
  (NMOS, $I_D = 20\,\mu$A).}\;
Identical bias point to $NM_2$; the Cadence sweep gives
$W = 3.99\,\mu\text{m}$, rounded to $4\,\mu$m.

\medskip
\noindent\textbf{$PM_2$, $PM_0$ --- PMOS current-mirror load
  ($I_D = 10\,\mu$A each).}\;
$PM_2$ is diode-connected with $V_S = 1.8\,\text{V}$ and
$V_G = V_D = 0.9\,\text{V}$.
The Cadence sweep gives
$W = 922.934\,\text{nm}$, rounded to $920$\,nm in the schematic.

\medskip
\noindent\textbf{$PM_1$ --- Common-source output stage
  (PMOS, $I_D = 20\,\mu$A).}\;
The Cadence sweep gives $W = 1.83\,\mu\text{m}$. A 200\,fF Miller capacitor
($C_2$) compensates the amplifier, and $C_1 = 2$\,pF loads the output.

\subsubsection{Total Quiescent Power Consumption}

The total quiescent current drawn by the complete BGR is estimated from the
DC operating point at 27\,\textdegree{}C. The opamp stages draw
$I_{op} = 20\,\mu$A (Stage 1 tail) $+\,20\,\mu$A (Stage 2) $= 40\,\mu$A.
Each of the two BGR core branches carries
$I_{C1} + V_{BE1}/R_3 \approx 6.2 + 0.61\,\text{V}/85\,\text{k}\Omega
\approx 13.4\,\mu$A, and the output branch carries
$V_{REF}/R_4 = 0.9\,\text{V}/69.1\,\text{k}\Omega \approx 13.0\,\mu$A, for a
core total of about 39.8\,$\mu$A. The start-up divider $R_0$--$R_1$ draws
$1.8\,\text{V}/180\,\text{k}\Omega = 10\,\mu$A. The total quiescent current is
therefore approximately $I_{DD,q} \approx 90\,\mu$A, giving a total static
power of:
\begin{equation}
  P_{q} = V_{DD} \times I_{DD,q} \approx 1.8\,\text{V} \times 90\,\mu\text{A}
        \approx 162\,\mu\text{W}.
  \label{eq:power}
\end{equation}
The opamp and the core branches dominate the consumption. Future work
targeting ultra-low-power applications may reduce $I_{DD,q}$ by lowering the
opamp bias currents and by replacing the resistive start-up divider.

\subsubsection{Open-Loop Gain Simulation}

\begin{figure}[H]
  \centering
    \includegraphics[width=\columnwidth]{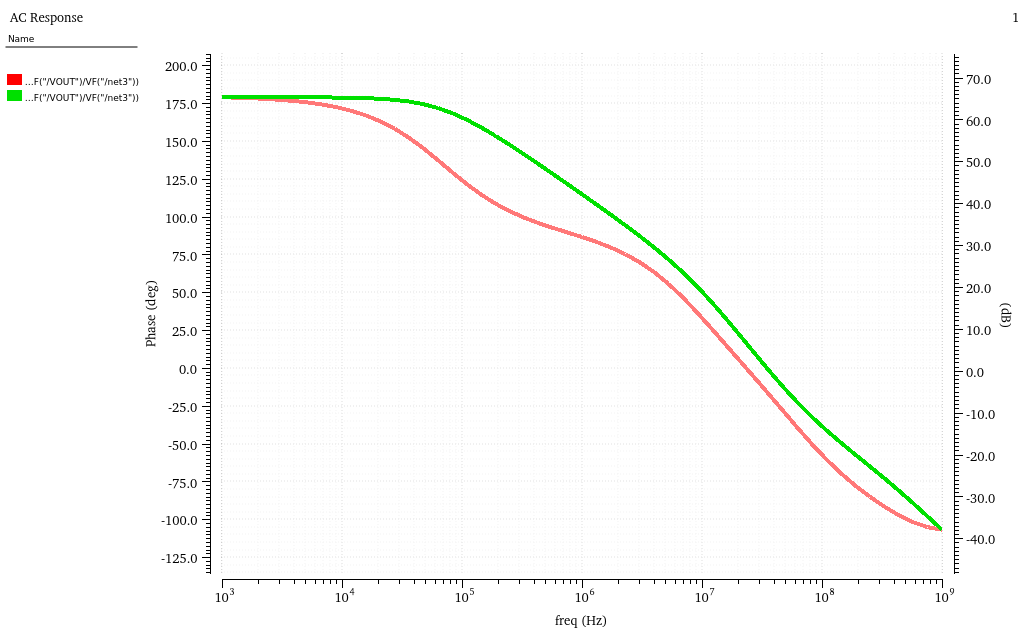}
  \caption{Simulated open-loop gain (dB, right axis) and phase (left axis)
           vs.\ frequency of the two-stage opamp.}
  \label{fig:opamp_gain}
\end{figure}

Figure~\ref{fig:opamp_gain} shows the AC open-loop response of the designed
opamp. The DC gain is approximately \textbf{65\,dB}, providing adequate loop
gain for the BGR feedback network to enforce $V_A = V_B$ with negligible
systematic offset; Fig.~\ref{fig:VA_VB} confirms that $V_A$ and $V_B$ remain
equal over the full temperature range of $-200$\,\textdegree{}C to
$+200$\,\textdegree{}C.

\subsection{Start-Up Circuit}

A bandgap voltage reference inherently possesses two stable operating
points: the desired equilibrium state, in which the bias network carries
the correct quiescent currents, and a degenerate zero-current state in
which $I_1 = I_2 = 0$ and $V_A = V_B = 0$\,V. Without intervention,
the feedback loop can lock into this zero-current condition at power-on,
permanently preventing the circuit from reaching its intended operating
point. A dedicated start-up circuit is therefore necessary to guarantee
that the BGR always converges to its correct bias state.

\begin{figure}[H]
  \centering
  \includegraphics[width=\columnwidth]{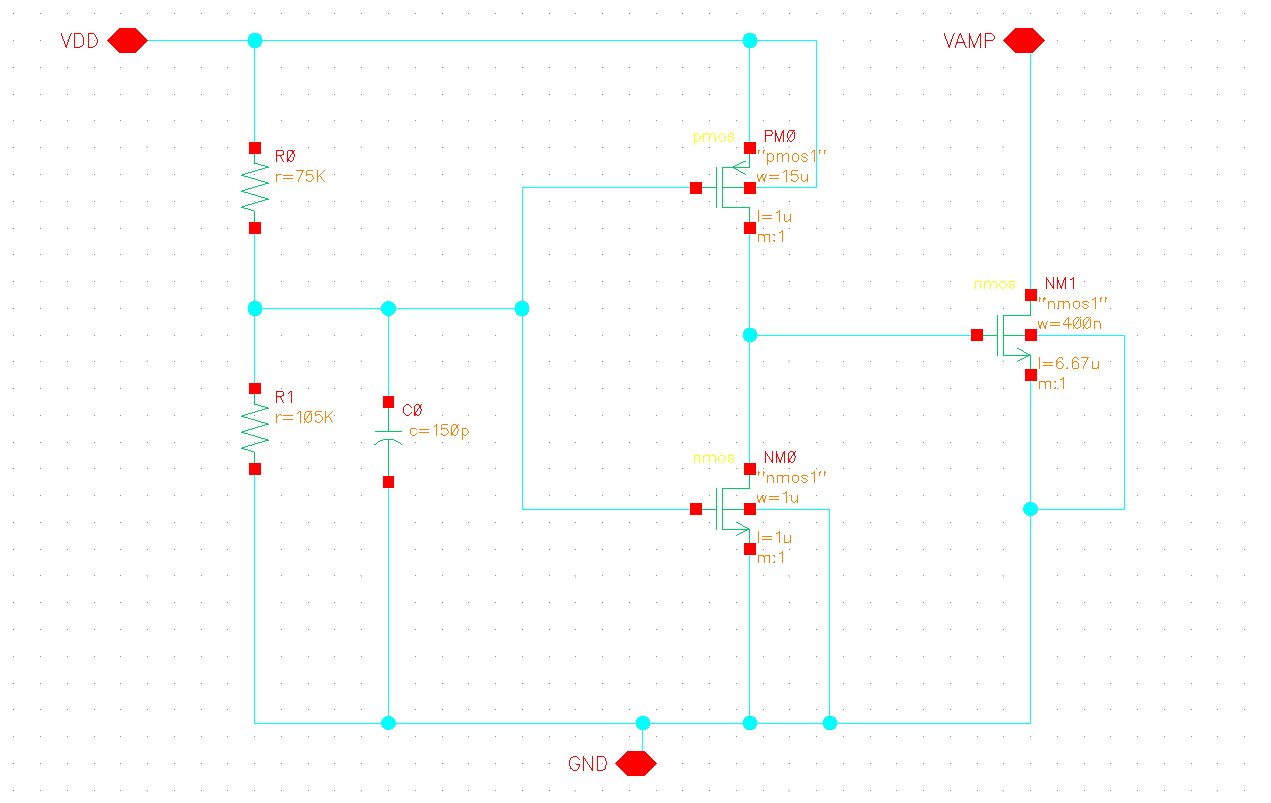}
  \caption{Start-up circuit schematic.}
  \label{fig:startup_sch}
\end{figure}

\subsubsection{Operating Principle}

The start-up circuit of Fig.~\ref{fig:startup_sch} exploits an RC time delay
to generate a transient control pulse at power-on. When $V_{DD}$ is first applied, capacitor
$C_0$ is initially uncharged, so the input of the CMOS inverter formed
by $NM_0$ and $PM_0$ is held near GND. The inverter output is therefore
high, turning on transistor $NM_1$. With $NM_1$ conducting, the gate
node \textit{VAMP}, connected to the gates of the PMOS mirror
transistors $M_1$ and $M_2$ in the BGR core, is pulled toward GND,
forcing a gate-to-source voltage sufficient to switch on the mirror and
initiate current flow in the BGR branches. This breaks the zero-current
lock and drives the circuit toward its correct equilibrium.
$C_0$ charges through the divider $R_0$--$R_1$ with time constant
\begin{equation}
  \tau = (R_{0} \parallel R_{1}) \times C_0
       = 43.75\;\text{k}\Omega \times 150\;\text{pF}
       \approx 6.56\;\mu\text{s},
  \label{eq:startup_tau}
\end{equation}
towards the final value $V_{DD}\,R_1/(R_0+R_1) = 1.05$\,V. Once the inverter
input crosses the inverter switching threshold, the inverter output
transitions low, turning off $NM_1$.
From this point the start-up path is inactive and presents negligible
loading to the BGR core. Resistor $R_{1}$ also discharges $C_0$ when
$V_{DD}$ is removed, resetting the circuit for the next power-on cycle;
$R_0$ and $R_1$ are chosen large ($75$ and $105\;\text{k}\Omega$) to limit
the continuous current through this path to 10\,$\mu$A during normal
operation.

\subsubsection{Component Sizing}
Tables~\ref{tab:startup_RC} and~\ref{tab:startup_mos} list the passive
and active component values used in the start-up circuit.

\begin{table}[H]
\centering
\caption{Start-Up Circuit: Passive Component Values}
\label{tab:startup_RC}
\renewcommand{\arraystretch}{1.3}
\begin{tabular}{lcc}
\toprule
\textbf{Component} & \textbf{Value} & \textbf{Unit} \\
\midrule
$R_{0}$ & 75  & k$\Omega$ \\
$R_{1}$ & 105 & k$\Omega$ \\
$C_0$   & 150 & pF        \\
\bottomrule
\end{tabular}
\end{table}

\vspace{4pt}

\begin{table}[H]
\centering
\caption{Start-Up Circuit: MOSFET Dimensions (GPDK 180)}
\label{tab:startup_mos}
\renewcommand{\arraystretch}{1.3}
\begin{tabular}{llcc}
\toprule
\textbf{Transistor} & \textbf{Type}
  & \textbf{W\,(\textmu m)} & \textbf{L\,(\textmu m)} \\
\midrule
$NM_0$ & NMOS & 1.0  & 1.0 \\
$NM_1$ & NMOS & 0.4  & 6.67 \\
$PM_0$ & PMOS & 15.0 & 1.0 \\
\bottomrule
\end{tabular}
\end{table}

\vspace{2pt}

$NM_1$ is sized with a long channel ($L = 6.67\,\mu$m) to ensure it remains
off reliably once the inverter transitions low, preventing any leakage
current from disturbing the BGR mirror gate after start-up. $PM_0$ is
sized wide ($W = 15\,\mu$m) to ensure the inverter switches sharply,
minimising the transition time and reducing uncertainty in the start-up
delay.

\begin{figure}[H]
  \centering
  \includegraphics[width=\columnwidth]{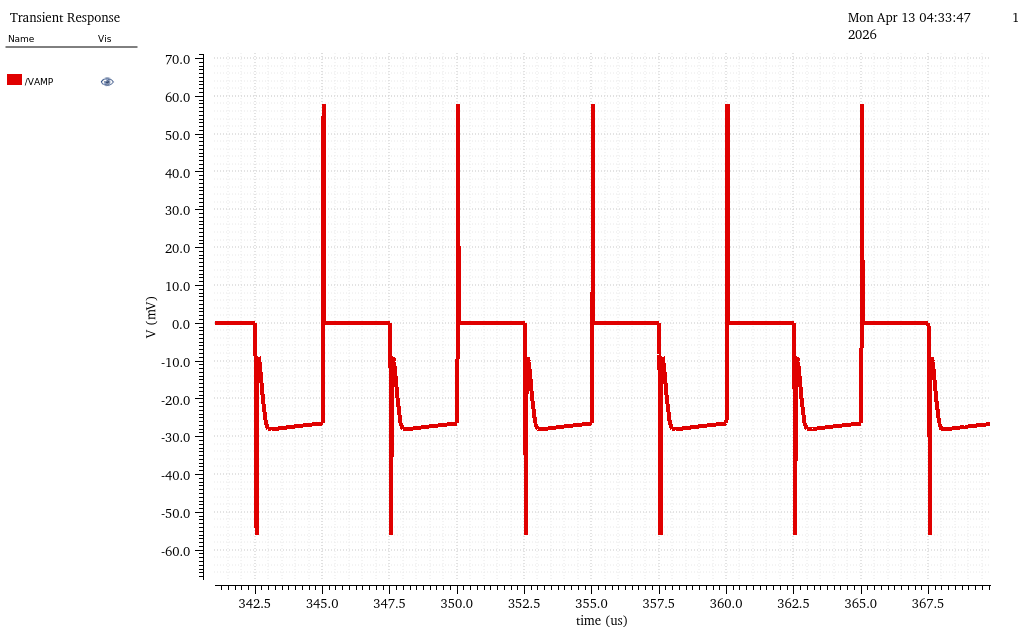}
  \caption{Simulated transient voltage at the \textit{VAMP} node of the
           start-up circuit (342--368\,$\mu$s window).}
  \label{fig:startup_tran}
\end{figure}

Figure~\ref{fig:startup_tran} shows the transient simulation of the
\textit{VAMP} node driven by the start-up circuit. Each time the start-up
path is triggered, $NM_1$ pulls \textit{VAMP} low by a few tens of millivolts,
which is the pull-down action that injects gate drive into the PMOS mirror,
after which the node is released and $NM_1$ turns off.

\begin{figure*}[!t]
  \centering
  \includegraphics[width=\textwidth]{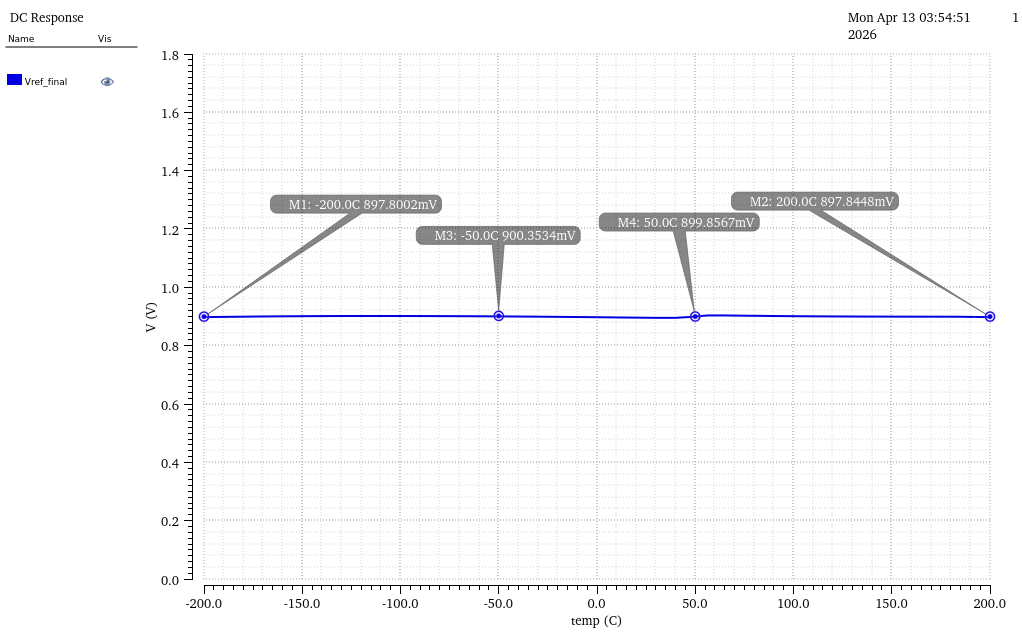}
  \caption{Simulated $V_{REF}$ vs.\ temperature over
           $-200$\,\textdegree{}C to $+200$\,\textdegree{}C.
           Key markers:
           $V_{REF}(-200\,\text{\textdegree{}C}) = 897.8002$\,mV,
           $V_{REF}(-50\,\text{\textdegree{}C})  = 900.3534$\,mV,
           $V_{REF}(+50\,\text{\textdegree{}C})  = 899.8567$\,mV,
           $V_{REF}(+200\,\text{\textdegree{}C}) = 897.8448$\,mV.}
  \label{fig:vref_temp}
\end{figure*}

\section{Simulation Results and Discussion}

\textbf{The proposed BGR achieves a temperature coefficient of
7.09\,ppm/\textdegree{}C over a 400\,\textdegree{}C span
($-200$\,\textdegree{}C to $+200$\,\textdegree{}C), a span two to four times
wider than that of the prior works in Table~\ref{tab:perf}.} All simulations
were performed in Cadence Virtuoso using the GPDK 180\,nm process design kit
at the typical process corner. Analyses include DC operating point,
temperature sweep, AC analysis, transient analysis and power consumption
estimation.

\subsection{Reference Voltage vs.\ Temperature}

Figure~\ref{fig:vref_temp} presents the DC temperature sweep of $V_{REF}$
across the full range of $-200$\,\textdegree{}C to $+200$\,\textdegree{}C.
The output exhibits the characteristic shallow curvature typical of a
first-order compensated BGR: the reference rises slightly from its
low-temperature value, reaches a broad peak in the vicinity of
$-50$\,\textdegree{}C to $+50$\,\textdegree{}C, and returns toward the
low-temperature value at the upper extreme. A small step is visible near
$+50$\,\textdegree{}C, which also appears in the collector currents of
Fig.~\ref{fig:IC_equal}. Across the 400\,\textdegree{}C span the marker
values remain within the narrow band of 897.8\,mV to 900.4\,mV, a total
variation of only 2.55\,mV, confirming that the PTAT and CTAT contributions
are well balanced by the resistor network.

The temperature coefficient is computed using the marker extremes
and the nominal output at room temperature ($V_{REF,nom} \approx 899.9$\,mV):
\begin{equation}
TC = \frac{V_{REF,max} - V_{REF,min}}
{(T_{max} - T_{min}) \times V_{REF,nom}} \times 10^{6},
\end{equation}
\begin{equation}
  TC = \frac{900.3534 - 897.8002}{899.9 \times 400}
       \times 10^{6}
     \approx \mathbf{7.09\;ppm/\textdegree{}C.}
  \label{eq:TC_calc}
\end{equation}

This TC is lower than the average TC reported in~\cite{leung2002} and
\cite{ueno2009} (15\,ppm/\textdegree{}C) and is obtained over a temperature
range four times wider, as summarised in Table~\ref{tab:perf}.

\subsection{Temperature Slope of \texorpdfstring{$V_{REF}$}{VREF}}

\begin{figure}[H]
  \centering
  \includegraphics[width=\columnwidth]{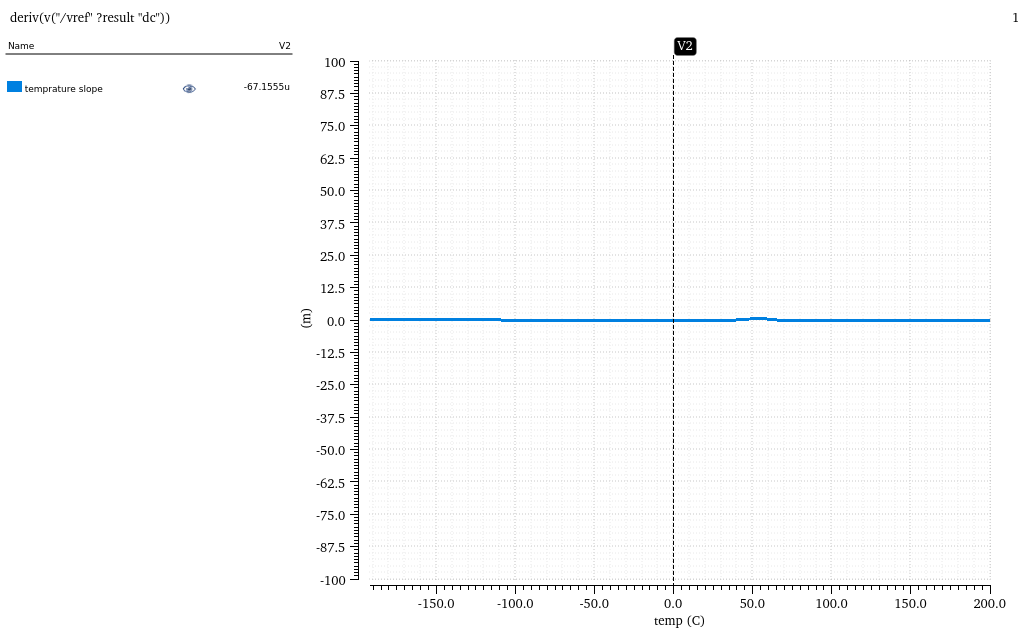}
  \caption{Simulated temperature derivative
           $\partial V_{REF}/\partial T$ vs.\ temperature (cursor at
           0\,\textdegree{}C: $-67.16\,\mu$V/\textdegree{}C).}
  \label{fig:vref_slope}
\end{figure}

Figure~\ref{fig:vref_slope} plots the first derivative
$\partial V_{REF}/\partial T$ across the full sweep. The derivative
waveform remains essentially flat and near zero across the entire temperature
axis, apart from a small excursion near $+50$\,\textdegree{}C. At
0\,\textdegree{}C the slope is $-67.16\,\mu$V/\textdegree{}C, about 27 times
smaller than the raw CTAT slope of $V_{BE}$
($\approx -1.84$\,mV/\textdegree{}C), confirming that the PTAT correction
provided by the resistor network is highly effective. The residual slope
arises from the second-order curvature in $V_{BE}(T)$ inherent to
first-order BGR topologies; higher-order compensation could reduce this further.

\subsection{Performance Summary}

Table~\ref{tab:perf} summarises the key performance metrics of the
proposed BGR and benchmarks them against representative prior works.

\begin{table}[H]
\centering
\caption{Performance Summary and Comparison with Prior Work}
\label{tab:perf}
\renewcommand{\arraystretch}{1.3}
\resizebox{\columnwidth}{!}{%
\begin{tabular}{lcccc}
\toprule
\textbf{Parameter}
  & \textbf{This Work}
  & \cite{banba1999}
  & \cite{leung2002}
  & \cite{ueno2009} \\
\midrule
Result type
  & Simulation & Measured & Measured & Measured \\
Technology (nm)
  & 180 & 400 & --- & 350 \\
Supply $V_{DD}$ (V)
  & 1.8 & --- & $\geq$0.98 & 1.4--3 \\
$V_{REF}$ (V)
  & 0.9 & 0.518 & --- & 0.745 \\
TC (ppm/\textdegree{}C)
  & \textbf{7.09} & --- & 15 & 15 (avg.) \\
Opamp Gain (dB)
  & 65 & --- & --- & --- \\
Power ($\mu$W)
  & $\approx$162 & --- & --- & 0.3 \\
\textbf{Temp.\ Range (\textdegree{}C)}
  & $\mathbf{-200\ \text{to}\ +200}$
  & $27$ to $125$
  & $0$ to $100$
  & $-20$ to $+80$ \\
\textbf{Temp.\ Span (\textdegree{}C)}
  & \textbf{400} & 98 & 100 & 100 \\
\bottomrule
\end{tabular}%
}
\end{table}

The proposed design achieves a TC of 7.09\,ppm/\textdegree{}C, lower than
the 15\,ppm/\textdegree{}C of~\cite{leung2002} and~\cite{ueno2009}, over a
temperature range four times wider. Curvature-compensated references reach
lower TCs, such as 3.9\,ppm/\textdegree{}C over 165\,\textdegree{}C
in~\cite{andreou2012} and 5.3\,ppm/\textdegree{}C in~\cite{leung2003}, at the
cost of additional circuitry. The small residual slope at 0\,\textdegree{}C
further validates the near-ideal cancellation achieved by the opamp-enforced
$V_A = V_B$ feedback condition and the precision resistor-ratio design. The
prior works are measured silicon, while the present results are simulations
at the typical corner; a fair comparison requires the silicon validation
planned as future work.

\section{Conclusion}

This work demonstrates that first-order TC compensation, when implemented
using a precision opamp-based $\beta$-multiplier topology with a carefully
ratioed BJT emitter-area and resistor network, can achieve
\textbf{7.09\,ppm/\textdegree{}C over a 400\,\textdegree{}C
range} ($-200$\,\textdegree{}C to $+200$\,\textdegree{}C) in simulation, a
temperature span two to four times wider than that of the sub-1\,V and
wide-range references compared. This level of thermal stability is
particularly significant for applications
in automotive electronics, avionics, deep-space instrumentation, and
industrial control systems, where reference circuits must maintain accuracy
across extreme temperature excursions without recalibration.
The circuit was implemented in GPDK 180\,nm CMOS technology producing
$V_{REF} \approx 0.9$\,V from a 1.8\,V supply. A custom two-stage
operational amplifier with 65\,dB open-loop gain accurately enforces
$V_A = V_B$ across the full temperature range. The BGR core was designed with
a bias of $I_{C1} = I_{C2} = 6.20196\,\mu$A at 27\,\textdegree{}C and resistor
network $R_1 = 8.671$\,k$\Omega$, $R_2 = R_3 \approx 85$\,k$\Omega$,
$R_4 = 69.1$\,k$\Omega$. The temperature derivative of only
$-67.16\,\mu$V/\textdegree{}C at 0\,\textdegree{}C, about 27 times below the
raw CTAT slope, confirms the effectiveness of the compensation scheme. Total
quiescent power consumption is estimated at about 162\,$\mu$W.

The results are simulations at the typical process corner; the device models
of a standard process kit are not normally characterised down to
$-200$\,\textdegree{}C, so the low-temperature end of the range should be
confirmed in silicon. Future work will address second-order curvature
correction to further reduce TC, process-corner and Monte Carlo analysis,
PSRR and noise characterisation, layout optimisation for improved matching,
and measurement of a fabricated prototype.

\FloatBarrier
\bibliographystyle{IEEEtran}
\bibliography{refs}

\end{document}